# *In operando* cryo-STEM of pulse-induced charge density wave switching in TaS$_2$

James L Hart[1], Saif Siddique[1], Noah Schnitzer[1], Stephen D. Funni[1], Lena F. Kourkoutis[2,3], Judy J. Cha[1]

1. Department of Materials Science and Engineering, Cornell University, United States
2. School of Applied and Engineering Physics, Cornell University, United States
3. Kavli Institute at Cornell for Nanoscale Science, Cornell University, United States
*Corresponding author: jc476@cornell.edu

The charge density wave (CDW) material 1T-TaS$_2$ exhibits a pulse-induced insulator-to-metal transition, which shows promise for next-generation electronics such as memristive memory and neuromorphic hardware. However, the rational design of TaS$_2$ devices is hindered by a poor understanding of the switching mechanism, the pulse-induced phase, and the influence of material defects. Here, we operate a 2-terminal TaS$_2$ device within a scanning transmission electron microscope (STEM) at cryogenic temperature, and directly visualize the changing CDW structure with nanoscale spatial resolution and down to 300 µs temporal resolution. We show that the pulse-induced transition is driven by Joule heating, and that the pulse-induced state corresponds to nearly commensurate and incommensurate CDW phases, depending on the applied voltage amplitude. With our *in operando* cryo-STEM experiments, we directly correlate the CDW structure with the device resistance, and show that dislocations significantly impact device performance. This work resolves fundamental questions of resistive switching in TaS$_2$ devices critical for engineering reliable and scalable TaS$_2$ electronics.

*Introduction:*

1T-TaS$_2$ is a layered, two-dimensional (2D) quantum material which undergoes an insulator-to-metal transition induced by voltage pulses (Fig. 1a,b)[1–4]. The switching is fast, energy efficient, and reversible, making TaS$_2$ attractive for device applications[5,6]. Moreover, the layered structure of TaS$_2$ may enable atomically-thin memristive or neuromorphic devices, providing ultimate scalability inaccessible to 3D crystals[7]. Nevertheless, knowledge of the switching mechanism in TaS$_2$ is limited. Prior works indicate that the electrical switching is associated with the CDW structure (Fig. 1c)[1–4,8–12]. At low-temperature (< 200 K), TaS$_2$ exhibits the commensurate (C) CDW phase, which is insulating[12–14]. At higher temperature, several metallic CDW phases exist[12–14], as well as non-thermal CDW phases accessible with optical excitation[15,16]. Direct characterization of the CDW structure during device operation is limited. *In operando* scanning tunneling microscopy studies have visualized the CDW structure before and after switching[11,17], but scanning probe studies do not possess the time-resolution to capture the switching process, and are strictly surface sensitive. TaS$_2$ switching has also been studied with *in operando* optical measurements[8], but such measurements lack nanoscale spatial resolution, and only indirectly probe the CDW phase. Hence, our understanding of the bias-induced phase(s) is still unclear, and more importantly, the switching mechanism remains unknown. Most studies argue that the transition is field-induced (non-thermal), although there are competing proposals for the microscopic mechanism[1–6,8]. More recently, several groups have claimed that Joule heating is partially or wholly responsible for switching based on finite element simulations[9,10], as well as IR thermal imaging of a bulk crystal under constant bias[18]. A concrete understanding of the switching mechanism is critical for the development of TaS$_2$ electronics. To this end, *in operando* measurements are needed to correlate the CDW structure, flake temperature, and electrical resistance of a nanoscale device.

Here, we operate a 2-terminal TaS$_2$ device within a scanning transmission electron microscope (STEM) at cryogenic temperature. Through time-resolved electron diffraction and 4D-STEM imaging, we quantify the CDW order parameter during electric biasing and, *via* strain analysis, we measure the local sample temperature. By directly correlating the CDW structure, flake temperature, and device resistance during switching, we unequivocally show that Joule heating drives the switching process, both for steady-state bias and short voltage pulses. Accordingly, the bias-induced phases correspond to thermal CDW states. We also show coupling between the CDW order parameter and the device resistance, and we demonstrate how local microstructural features (dislocations) influence device operation. These findings are crucial for the engineering and optimization of TaS$_2$ devices for beyond-silicon technology.

*Results and Discussion:*

The studied device is shown in Fig. 1d and 1e with optical and STEM imaging, respectively. A bulk 1T-TaS$_2$ crystal was exfoliated in an Argon glove box onto a SiO$_2$ / Si substrate, and graphite electrodes were placed on the TaS$_2$ flake (channel length = 7 μm). The finished device was transferred to an *in situ* TEM chip, and placed over a through-hole drilled in an amorphous SiN$_x$ membrane. The TEM chip has Pt electrodes for *in situ* electric biasing, as well as a Pt coil that allows local sample heating and thermometry from ~100 – 1000 K[19,20].

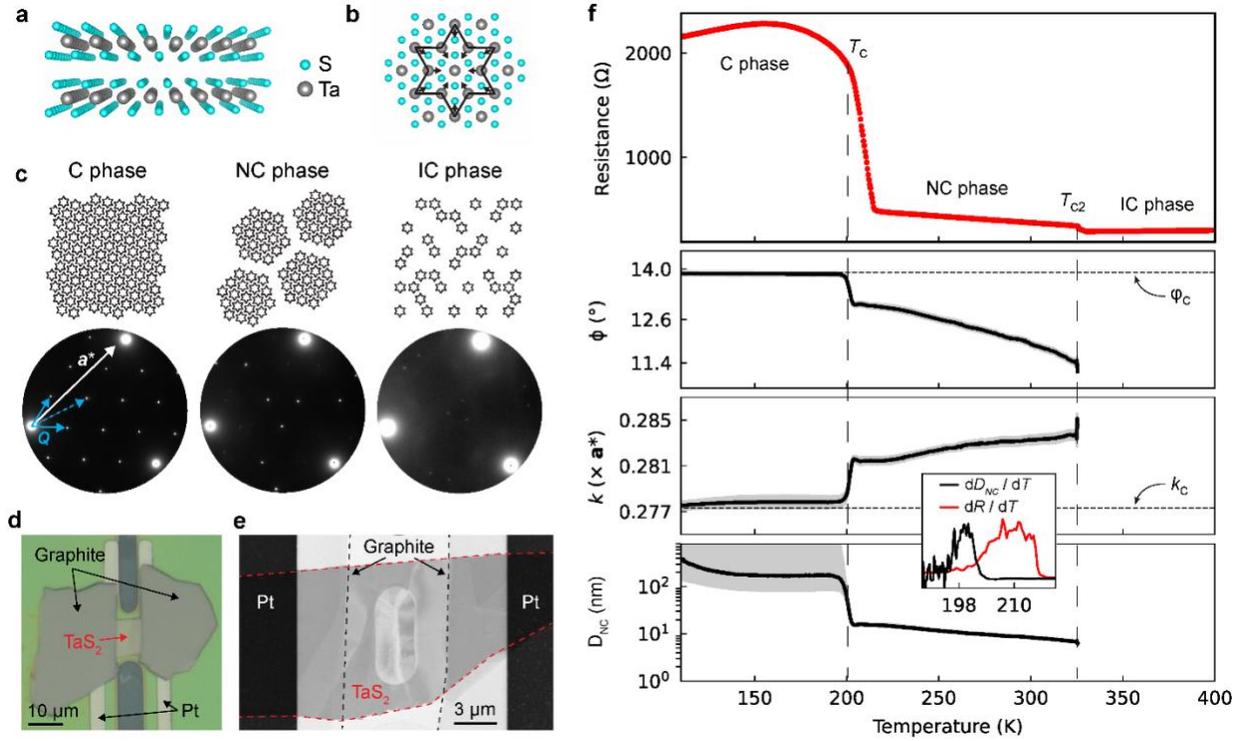

**Figure 1 | TaS$_2$ structure and temperature dependent CDW behavior.** Atomic structure of 1T-TaS$_2$ in cross-section (**a**) and plan-view (**b**). In **b**, the local CDW distortion is shown, which forms a Star-of-David structure. **c**. Illustration of the C, NC, and IC CDW phases, which exhibit different orderings of the stars. Associated electron diffraction patterns are shown. For the C phase diffraction pattern, the Bragg vector ***a*** * is shown, as well as two first order CDW ***Q*** vectors (solid line) and one 2$^{nd}$ order CDW vector (dotted line). The studied device imaged optically (**d**) and with STEM high angle annular dark field imaging (**e**). In the center of the STEM image is a through-hole in the SiN$_x$ membrane, which allows for electron diffraction measurements. **f**. Temperature dependence of the TaS$_2$ resistance, CDW angle φ, CDW wavevector magnitude $k$, and the domain size D$_{NC}$. The shaded regions represent the standard error. The inset shows the temperature derivatives of the resistance and D$_{NC}$. For the inset, the x-axis units are temperature (K). The temperature-dependent diffraction data is shown in Supplementary Video 1.

We use electron diffraction to characterize the CDW state. The TaS$_2$ CDW follows the Star-of-David distortion, wherein 13 Ta atoms bunch together (Fig. 1b). In the insulating low-temperature C phase, these stars form a long-range lattice commensurate with the atomic structure[13,14]. As shown in Fig. 1c, electron diffraction of the C phase yields sharp 1$^{st}$ and 2$^{nd}$ order CDW satellite peaks. The C phase CDW wavevector has an angle of φ$_c$ = 13.9° relative to the Bragg wavevector ***a*** * and a magnitude of $k_c$ = 0.2773a*. Above ~200 K, defects in CDW known as discommensurations form and organize into a hexagonal network with a CDW domain size of order ~10 nm[13,14]. This is the metallic nearly commensurate (NC) phase, which suppresses the 1$^{st}$ order CDW peak intensity, and slightly adjusts the CDW wavevector, with φ ~ 11 – 13° and $k$ ~ 0.280a* – 0.285a*. Thomson *et al* showed that the values of φ and $k$ determine the domain size of the NC phase, D$_{NC}$, according to

$$D_{NC} = a/\sqrt{(2\pi\Delta\varphi/360)^2 + (\Delta k/k_C)^2} \qquad \text{Equation 1}$$

where $a$ is the atomic lattice parameter, and $\Delta\varphi$ and $\Delta k$ are the differences in $\varphi$ and $k$ relative to their commensurate values[14]. Hence, $D_{NC}$ provides an appropriate order parameter to differentiate the C and NC phases, with $D_{NC} = \infty$ for the C phase. Above ~325 K, TaS$_2$ transitions to the incommensurate (IC) phase, with $\varphi \sim 0°$ and $k \sim 0.285a*$, and Equation 1 is no longer applicable.

We first study the CDW behavior as a function of temperature. Figure 1f shows the flake resistance upon heating at 0.4 K / s; the C to NC and NC to IC transitions are clearly observed. In total, the resistance drops by a factor of 8. Using a direct detection camera[21], we collect temperature-resolved selected area electron diffraction data simultaneous with the resistance measurements (Supplementary Video 1). From the diffraction data, we quantify the CDW structure based on the 2$^{nd}$ order CDW spots (Supplementary Note 1 and Supplementary Figure 1). The CDW angle $\varphi$ and magnitude $k$ are plotted in Fig. 1f, as well as the calculated $D_{NC}$. In the low-temperature C phase, $D_{NC} \sim 500$ nm. The large error bars in the C phase reflect the nature of Equation 1; small errors in $\varphi$ and $k$ are magnified in the propagated $D_{NC}$ error as $\Delta\varphi$ and $\Delta k \rightarrow 0$. Upon entering the NC phase at ~ 200 K, $D_{NC}$ quickly falls to ~12 nm, and then gradually decreases to 8 nm before the flake enters the IC phase. By plotting the derivates $dD_{NC} / dT$ and $dR / dT$ (inset), we see that the structural CDW transition precedes the resistive transition by ~10 K. This finding was only possible given our *in operando* multimodal experimental approach. As we show later, this result is relevant to device operation.

Next, we study bias-induced CDW switching using triangular voltage ramps with durations of 20 s (Fig. 2a inset), with diffraction patterns collected at a rate of 100 Hz. These slow ramps effectively probe the steady-state CDW response to an applied electric field. Measurements are performed at ~110 K, starting in the C phase. Applying triangular voltage ramps with maximum voltages ranging from 0.1 to 0.7 V yields triangular current vs time curves, perfectly reflecting the input voltage profile and indicating no CDW switching (Fig. 2a). For this voltage range, there are minimal changes in the CDW structure, as seen in the domain size $D_{NC}$ data (Fig. 2b and Supplementary Video 2). In contrast, for voltages above 0.8 V, there is a sudden increase in current, indicating resistive switching. Concurrently, $D_{NC}$ rapidly falls, indicating the C to NC transition, quickly followed by the NC to IC phase transition (Fig. 2b, insets, Supplementary Figure 2, and Supplementary Video 3). As the applied voltage decreases on the second half of the voltage ramp, the flake recovers to the NC phase, and then slowly progresses towards the C phase over the next several minutes. For this device, the steady-state switching threshold is ~ 0.8 V. While the measured $I$ vs $V$ behavior is consistent with data in the literature[1–4], this experiment constitutes the first *in operando* study to directly quantify the CDW order parameter during switching, and to identify the NC and IC phases as the bias-induced states.

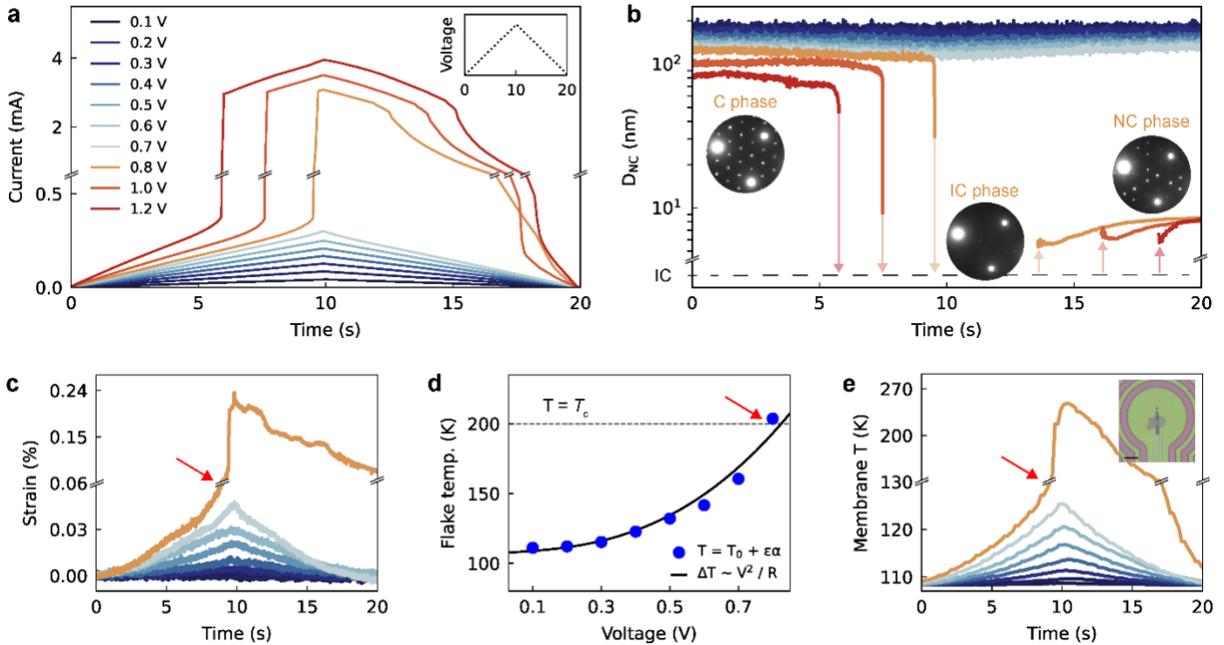

**Figure 2 | Steady-state biasing and CDW switching. A.** Current vs time during triangular voltage ramps with the maximum voltage ranging from 0.1 V to 1.2 V. The maximum voltage is reached at 10 s. Inset: example voltage profile, the *x*-axis is time (s). The color legend in **a** applies to **b**, **c,** and **e** as well. **B.** The measured CDW domain size $D_{NC}$ during the voltage ramps. The insets show diffraction snapshots acquired during the 0.8 V ramp. The full diffraction datasets for 0.7 V and 0.8 V are shown in Supplementary Videos 2 and 3. **C.** Flake strain during the voltage ramps. **D.** Maximum flake temperature for voltage ramps from 0.1 to 0.8 V, calculated from the strain shown in **c**. For the 0.8 V datapoint, we show the temperature immediately prior to the C to NC transition. $T_0$ is 110 K, $\varepsilon$ is the strain, $\alpha$ is the effective coefficient of thermal expansion, and R is the flake resistance. **E.** Measured temperature of the SiN$_x$ membrane during voltage ramps. The thermometer consists of a Pt coil encompassing the flake, pictured in the inset. Scale bar is 50 μm.

We now demonstrate that the bias-induced switching is driven by Joule heating. To measure the local flake temperature during bias, we extract the in-plane flake strain ε from the diffraction data (Fig. 2c, Supplementary Note 2). For the sub-threshold voltage pulses of ≤ 0.7 V, the strain versus time profiles rise and fall with the applied voltage ramp, with larger voltages leading to increased strains. For the 0.8 V pulse, there is a sudden increase in strain at the bias-induced CDW transition. Next, we convert the strain data to temperature using the effective thermal coefficient of expansion α for this device (Supplemental Note 2)[22,23]. Figure 2d plots the maximum flake temperature during each voltage ramp based on the strain data shown in Fig. 2c. For the 0.8 V curve, we plot the temperature immediately prior to the CDW transition. The data clearly shows that with biasing at 0.8 V, the flake temperature surpasses 200 K, which is the $T_c$ for the C to NC phase transition (Fig. 1). Red arrows mark $T_{flake} \sim 200$ K in Figs. 2c and 2d. Thus, our data clearly shows that at the threshold voltage of 0.8 V, Joule heating is sufficient to raise the flake temperature to 200 K and thermally trigger the C to NC transition. Supporting this claim, the temperature versus voltage data is well fit by $\Delta T \sim P = V^2 / R$, where P is the power generated from Joule heating, V is the maximum applied voltage, and R is the flake resistance prior to switching. Joule heating also explains the rapid rise in strain (temperature) after switching at 0.8 V: as the CDW transition begins and the resistance drops, the power generated increases by 1 / R, which further increases the flake temperature and accelerates the transition. This positive feedback loop leads to sudden and complete CDW switching, and, consequently, a spike in the flake temperature and strain. Additionally, there is an inherent lattice expansion at the C to NC transition[22,24], which contributes to the strain jump at the CDW transition. This effect also results in a remnant positive strain after the voltage ramp is complete. Because the CDW structure does not fully relax to the C phase within the measurement time-frame, the strain remains finite due to the CDW-lattice coupling (Fig. 2c, 0.8 V curve).

To provide an independent confirmation of Joule heating, we plot the membrane temperature during the voltage ramps, as measured with the Pt coil on the TEM chip (Fig. 2e). The membrane temperature shows the same qualitative behavior as the flake strain and temperature measurements (Fig. 2c and 2d), supporting the presence of Joule heating. For the 0.8 V ramp immediately prior to the CDW transition, the membrane temperature is ~130 K. Using a simple thermal transport model, we find that a membrane temperature of 130 K is consistent with a flake temperature of 200 K (Supplementary Note 3). We also highlight that after the CDW transition, the membrane temperature shows a temperature spike of > 100 K, which supports the positive feedback loop between Joule heating and the insulator-to-metal transition.

It follows from the Joule heating hypothesis that the time needed for switching $t_s$ should scale inversely with the applied voltage[25,26]. Specifically, a model from Fangohr et al[27] predicts $t_s \sim \sinh^2(V^{-2})$ for a nanodevice on a 2D membrane. To evaluate this behavior, we measure the time-resolved resistance during switching using the biasing setup shown in Fig. 3a. We pass square voltage pulses (durations from 3 ms down to 100 μs) through a 1 kΩ resistor in series with the flake, and we plot the voltage drop across the flake divided by the total applied voltage, $V_{flake} / V_{total}$. This ratio scales with the flake resistance; hence, a drop in $V_{flake} / V_{total}$ indicates bias-induced switching (Fig. 3b). There is no switching for a 3 ms 1.4 V pulse, but all pulses > 1.4 V initiate switching, with progressively shorter switching times for larger voltage amplitudes. *In operando* diffraction measurements show switching from the C to the NC phase (Fig. 3c and Supplementary Video 4). Moreover, the time-resolved diffraction shows that an increase in flake strain (temperature) precedes the CDW transition, as expected from Joule heating (Supplementary Figure 4). The Fig. 3b inset plots $t_s$ as a function of $V_{total}$, and the $t_s \sim \sinh^2(V^{-2})$ model provides an excellent fit to the data. This result further confirms the Joule heating-induced switching mechanism. Conversely, for a non-thermal voltage induced mechanism, one would expect rapid (~ps) switching for all voltages above the threshold value[5,6], in clear contrast with our results. The $\sinh^2(V^{-2})$ fit also suggests that Joule heating can induce switching on the ns timescale, given a sufficiently large voltage pulse. Indeed, many literature reports of pulse-induced switching can reasonably be accounted for *via* Joule heating[1–4].

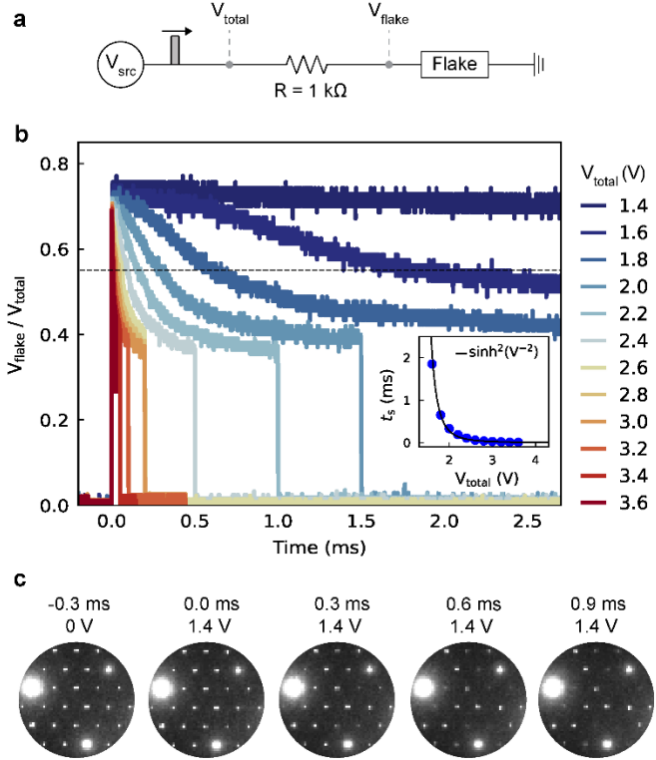

**Figure 3 | Time-dependent resistive switching. a.** Our biasing setup, where $V_{src}$ is the voltage source used to generate square voltage pulses, and $V_{total}$ and $V_{flake}$ are measured with an oscilloscope. **b.** The measured ratio of $V_{flake} / V_{total}$ for $V_{total}$ ranging from 1.4 up to 3.6 V. We define the switching time $t_s$ when $V_{flake} / V_{total}$ drops below 0.55, as shown with the horizontal dotted line. The inset plots $t_s$ versus $V_{total}$, along with a fit for Joule heating on a 2D membrane provided in ref. 27. **c.** *In operando* diffraction snapshots from the 1.4 V pulse. The voltage pulse is applied at a time of 0 ms, although there is an uncertainty of ~0.3 ms in the diffraction pattern timestamps. The diffraction data for this pulse is shown in Supplementary Video 4, and further analysis is provided in Supplementary Figure 4.

We next study coupling between the CDW order parameter and the device resistance through a series of short voltage pulses which are relevant for device operation. We apply pulses with $V_{total}$ starting at 2.0 V and increasing by 0.4 V up to 9.6 V, all with a 3 μs pulse duration, performed at 110 K. Figure 4a shows the measured $V_{flake} / V_{total}$ for a representative set of pulses. Note that the *RC* time constant for this device is τ ~ 460 ns (likely due to poor impedance matching throughout the *in situ* TEM set up), which places an upper limit on device operation speed. Partial switching is observed for $V_{total} \geq 3.2$ V, and full switching is observed for $V_{total} \geq 5$ V (Supplementary Video 5 for $V_{total} = 9.6$ V). This behavior is consistent with Joule heating and the $t_s \sim \sinh^2(V^{-2})$ model, which predicts $t_s = 1.1$ μs for $V_{total} = 5.2$ V.

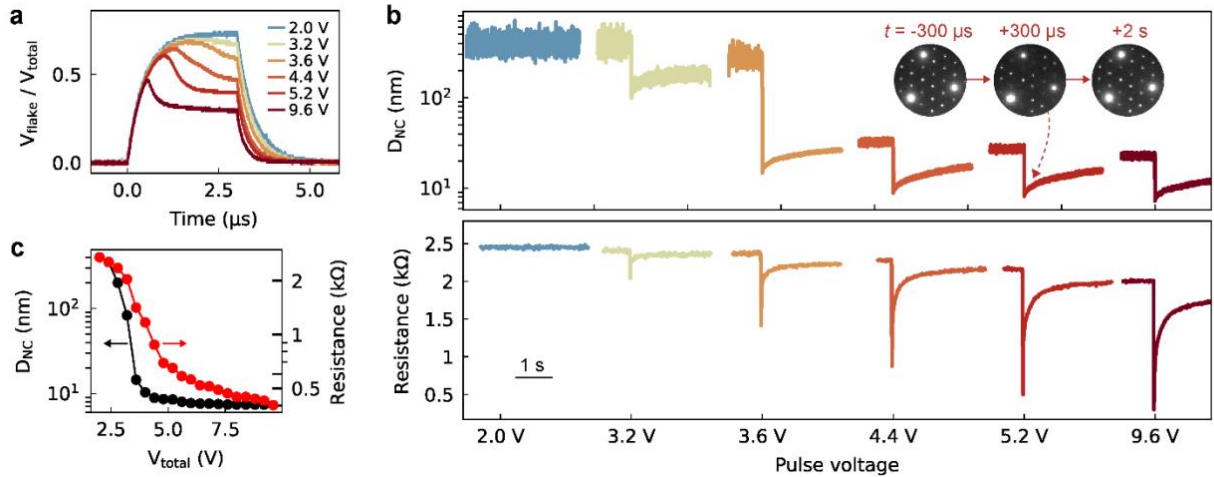

**Figure 4 | Pulse induced CDW and resistive switching. a.** $V_{total}$ / $V_{flake}$ during electric pulsing. Twenty consecutive pulses were performed in total, starting at 2.0 V and increasing by 0.4 V up to 9.6 V, with roughly 5 mins recovery time in between pulses. A representative set is shown here. **b.** Time-resolved CDW domain size $D_{NC}$ and device resistance during pulsing. The scale bar shows 1 second. The measurement time resolutions are 300 μs for the CDW analysis and 12 ms for the device resistance. The diffraction data for the 9.6 V pulse is shown in Supplementary Video 5. **c.** Comparison of the CDW domain size $D_{NC}$ (black) and the flake resistance (red) immediately after pulsing.

The pulse-dependent evolution of the CDW structure and device resistance is shown in Fig. 4b. Several interesting trends are present in the data. First, as the pulse amplitude increases, the pulse-induced $D_{NC}$ decreases, as does the device resistance. Thus, higher voltage pulses produce smaller CDW domains (thus a higher density of discommensurations), which in turn reduce the flake resistance. This behavior is captured in Fig. 4c, which plots the $D_{NC}$ and the flake resistance immediately after pulsing. This finding is consistent with Joule heating, with higher voltage pulses producing larger temperature changes. Comparing the $D_{NC}$ vs $V_{total}$ and resistance vs $V_{total}$ curves in Fig. 4c, we see that the structural CDW transition proceeds the electronic resistance transition. This trend is consistent with our earlier finding, that with heating through the C to NC transition, the $D_{NC}$ transition precedes the insulator-to-metal transition (Fig. 1f inset). Secondly, we find that after switching the device recovery is not complete, *i.e.* the switching is not fully reversible back to the C phase. This is evident in Fig. 4b; both the pre-pulse $D_{NC}$ and the device resistance steadily decrease after successive pulsing. We note that after the full pulsing experiment, the CDW structure and resistance were fully reset *via* heating to the IC phase and then cooling to 110 K.

The partial irreversibility of pulse-induced switching suggests CDW pinning on local microstructural features. Indeed, there are many TaS$_2$ biasing experiments which suggest increased CDW pinning in thin flakes[3,4,8], although direct confirmation is lacking. To test this hypothesis, we performed real-space analysis using 4D-STEM imaging. With this method, the electron beam is focused to a nanoscale probe, rastered across the sample surface, and a full diffraction pattern is captured at each spatial coordinate[28]. In turn, the $D_{NC}$ can be mapped in real space with a spatial resolution of ~15 nm. For these measurements we study a separate flake, imaged optically in the Fig. 5a inset. While the flake is a high-quality single crystal, we observe the presence of basal dislocations[29–31], which are revealed with a virtual-STEM image based on Bragg diffraction contrast (see the dark lines in Fig. 5a). We find that all exfoliated TaS$_2$ flakes (and many exfoliated

2D materials in general) exhibit similar dislocation structures. Figure 5b maps $D_{NC}$ as a function of constant applied bias. Initially, the $D_{NC}$ is mostly > 50 nm, indicating a spatially homogenous C phase, and the device is in high resistance state (resistance = 2.11 kΩ). With application of 0.5 V, the CDW remains in the C phase, and the device resistance remains high (1.99 kΩ). At 0.6 V, the device switches; the CDW map shows $D_{NC} \leq 10$ nm, and the resistance drops to 464 Ω. Note that this behavior is similar to our results in Fig. 2, albeit with a slightly lower threshold voltage, and the NC phase is more stable (for this flake, 0.8 V causes a transition to the IC phase). After releasing the applied bias, the sample resistance increases to 1.8 kΩ, and the CDW map mostly shows $D_{NC} > 50$ nm. However, the NC phase is found to locally persist at the basal dislocations, with $D_{NC} \sim 30$ nm adjacent to the line defects. This data suggests that after pulsing, discommensurations of the NC phase are pinned to dislocations, preventing complete recovery to the C phase. This real-space analysis provides a microscopic understanding of CDW pinning in $TaS_2$ devices.

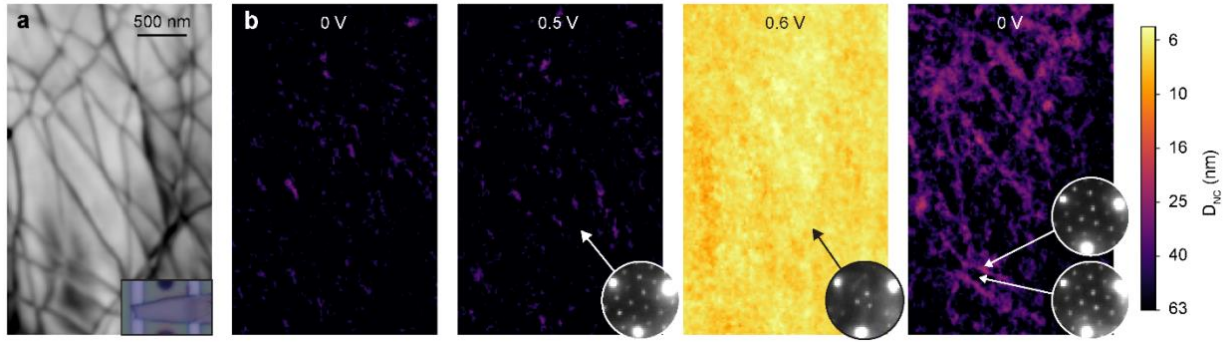

**Figure 5 | Real-space CDW imaging during bias. a.** Virtual STEM image which sums all of the Bragg peak intensities. The dark lines are basal dislocations. The inset is an optical image of this flake. **b.** Maps of the CDW $D_{NC}$ as a function of applied bias. The insets show cropped diffraction patterns, extracted from local regions of 3 x 3 pixels. For the post-bias dataset, the top diffraction pattern is extracted from a dislocation, and the bottom diffraction pattern is extracted from a non-defective region.

*Conclusion*:

By operating a 2-terminal $TaS_2$ device within the STEM at cryogenic temperature, we demonstrate that bias-induced switching is driven by Joule heating and a rapid thermal transition from the C to the NC and IC phases. In our device, this mechanism is operative for both steady-state biasing and μs voltage pulses. Our data also indicates that Joule heating can drive switching on the ns timescale. Based on this knowledge, we suggest engineering heat management of $TaS_2$ devices, *e.g.* the thermal conductivity of the substrate and the electrode geometry, in order to efficiently reach $T_c$ with minimal losses. While most of reports of $TaS_2$ switching are consistent with Joule heating, we note recent claims of picosecond switching, with a switching energy seemingly below the heat needed for Joule heating[25,26]. Hence, it may be that under certain conditions, purely field-induced switching is possible. Our finding that dislocations can pin the CDW structure and prevent complete recovery is relevant to the long-term reliability of $TaS_2$ devices under continuous operation. Devices could also be engineered with optimized dislocation structures to help stabilize the NC phase after switching, which could prolong the lifetime of the pulse-induced low-resistance state. Dislocation engineering may also enable faster switching as well as multi-level resistance states. Lastly, our use of *in operando* biasing and cryogenic cooling, along with novel 4D-STEM imaging of the CDW order parameter, offers a promising route to understand device performance for other quantum materials[32].

*References*:

1. Hollander, M. J. *et al.* Electrically Driven Reversible Insulator–Metal Phase Transition in 1T-TaS$_2$. *Nano Lett.* **15**, 1861–1866 (2015).
2. Vaskivskyi, I. *et al.* Fast electronic resistance switching involving hidden charge density wave states. *Nat. Commun.* **7**, 11442 (2016).
3. Tsen, A. W. *et al.* Structure and control of charge density waves in two-dimensional 1T-TaS$_2$. *Proc. Natl. Acad. Sci.* **112**, 15054–15059 (2015).
4. Yoshida, M., Suzuki, R., Zhang, Y., Nakano, M. & Iwasa, Y. Memristive phase switching in two-dimensional 1T-TaS$_2$ crystals. *Sci. Adv.* **1**, 1–7 (2015).
5. Mihailovic, D. *et al.* Ultrafast non-thermal and thermal switching in charge configuration memory devices based on 1T-TaS$_2$. *Appl. Phys. Lett.* **119**, (2021).
6. Venturini, R. *et al.* Ultraefficient resistance switching between charge ordered phases in 1 T -TaS$_2$ with a single picosecond electrical pulse. *Appl. Phys. Lett.* **120**, (2022).
7. Liu, X. & Hersam, M. C. 2D materials for quantum information science. *Nat. Rev. Mater.* **4**, 669–684 (2019).
8. Patel, T. *et al.* Photocurrent Imaging of Multi-Memristive Charge Density Wave Switching in Two-Dimensional 1T-TaS$_2$. *Nano Lett.* **20**, 7200–7206 (2020).
9. Geremew, A. K. *et al.* Bias-Voltage Driven Switching of the Charge-Density-Wave and Normal Metallic Phases in 1T-TaS$_2$ Thin-Film Devices. *ACS Nano* **13**, 7231–7240 (2019).
10. Mohammadzadeh, A. *et al.* Evidence for a thermally driven charge-density-wave transition in 1T-TaS$_2$ thin-film devices: Prospects for GHz switching speed. *Appl. Phys. Lett.* **118**, 093102 (2021).
11. Walker, S. M. *et al.* Observation and Manipulation of a Phase Separated State in a Charge Density Wave Material. *Nano Lett.* **22**, 1929–1936 (2022).
12. Wilson, J., Di Salvo, F. & Mahajan, S. Charge-density waves and superlattices in the metallic layered transition metal dichalcogenides. *Adv. Phys.* **24**, 117–201 (1975).
13. Ishiguro, T. & Sato, H. Electron microscopy of phase transformations in 1T - TaS$_2$. *Phys. Rev. B* **44**, 2046–2060 (1991).
14. Thomson, R. E., Burk, B., Zettl, A. & Clarke, J. Scanning tunneling microscopy of the charge-density-wave structure in 1T-TaS$_2$. *Phys. Rev. B* **49**, 16899–16916 (1994).
15. Gerasimenko, Y. A. *et al.* Quantum jamming transition to a correlated electron glass in 1T-TaS$_2$. *Nat. Mater.* **18**, 1078–1083 (2019).
16. Domröse, T. *et al.* Light-induced hexatic state in a layered quantum material. *Nat. Mater.* (2023) doi:10.1038/s41563-023-01600-6.
17. Mraz, A. *et al.* Energy efficient manipulation of topologically protected states in non-volatile ultrafast charge configuration memory devices. *arXiv* 1–19 (2021).
18. Jarach, Y., Rodes, L., Ber, E., Yalon, E. & Kanigel, A. Joule-heating induced phase transition in 1T-TaS$_2$ near room temperature probed by thermal imaging of power dissipation. *Appl. Phys. Lett.* **120**, 083502 (2022).
19. Goodge, B. H., Bianco, E., Schnitzer, N., Zandbergen, H. W. & Kourkoutis, L. F. Atomic-Resolution Cryo-STEM Across Continuously Variable Temperatures. *Microsc.*

*Microanal.* **26**, 439–446 (2020).
20. Garza, H. H. P. *et al.* MEMS-based sample carriers for simultaneous heating and biasing experiments: A platform for in-situ TEM analysis. in *2017 19th International Conference on Solid-State Sensors, Actuators and Microsystems (TRANSDUCERS)* 2155–2158 (IEEE, 2017). doi:10.1109/TRANSDUCERS.2017.7994502.
21. Philipp, H. T. *et al.* Very-High Dynamic Range, 10,000 Frames/Second Pixel Array Detector for Electron Microscopy. *Microsc. Microanal.* **28**, 425–440 (2022).
22. Sezerman, O., Simpson, A. M. & Jericho, M. H. Thermal expansion of lT-TaS$_2$ and 2H-NbSe$_2$. *Solid State Commun.* **36**, 737–740 (1980).
23. Givens, F. L. & Fredericks, G. E. Thermal expansion op NbSe$_2$ and TaS$_2$. *J. Phys. Chem. Solids* **38**, 1363–1365 (1977).
24. Philip, S. S., Neuefeind, J. C., Stone, M. B. & Louca, D. Local structure anomaly with the charge ordering transition of 1T-TaS$_2$. *Phys. Rev. B* **107**, 184109 (2023).
25. Liao, G. M. *et al.* Dynamically tracking the joule heating effect on the voltage induced metal-insulator transition in VO$_2$ crystal film. *AIP Adv.* **6**, (2016).
26. Li, D. *et al.* Joule Heating-Induced Metal–Insulator Transition in Epitaxial VO$_2$ / TiO$_2$ Devices. *ACS Appl. Mater. Interfaces* **8**, 12908–12914 (2016).
27. Fangohr, H., Chernyshenko, D. S., Franchin, M., Fischbacher, T. & Meier, G. Joule heating in nanowires. *Phys. Rev. B* **84**, 054437 (2011).
28. Ophus, C. Four-Dimensional Scanning Transmission Electron Microscopy (4D-STEM): From Scanning Nanodiffraction to Ptychography and Beyond. *Microsc. Microanal.* **25**, 563–582 (2019).
29. Medlin, D. L., Yang, N., Spataru, C. D., Hale, L. M. & Mishin, Y. Unraveling the dislocation core structure at a van der Waals gap in bismuth telluride. *Nat. Commun.* **10**, 1820 (2019).
30. Hovden, R. *et al.* Atomic lattice disorder in charge-density-wave phases of exfoliated dichalcogenides (1T-TaS$_2$). *Proc. Natl. Acad. Sci. U. S. A.* **113**, 11420–11424 (2016).
31. Hovden, R. *et al.* Thickness and Stacking Sequence Determination of Exfoliated Dichalcogenides (1T-TaS$_2$ , 2H-MoS$_2$ ) Using Scanning Transmission Electron Microscopy. *Microsc. Microanal.* **24**, 387–395 (2018).
32. Hart, J. L. & Cha, J. J. Seeing Quantum Materials with Cryogenic Transmission Electron Microscopy. *Nano Lett.* **21**, 5449–5452 (2021).
33. Sikora, A. *et al.* Highly sensitive thermal conductivity measurements of suspended membranes (SiN and diamond) using a 3ω-Völklein method. *Rev. Sci. Instrum.* **83**, (2012).